\documentclass[prd,aps,draft,showpacs,tightenlines,preprintnumbers]{revtex4}

\def\del{{\partial}}

\newcommand{\mathbb}[1]{\mbox{\Bbb #1}}

\def\lbldef#1#2{\expandafter\gdef\csname #1\endcsname {#2}}

\def\href#1#2{#2}

\newcommand{\beq}{\begin{equation}}
\newcommand{\eeq}{\end{equation}}
\newcommand{\bea}{\begin{eqnarray}}
\newcommand{\eea}{\end{eqnarray}}
\newcommand{\beqar}{\begin{eqnarray}}

\newcommand{\eeqar}{\end{eqnarray}}

\def\Box{\kern1pt\vbox{\hrule height 1.2pt\hbox{\vrule width 1.2pt\hskip 3pt
   \vbox{\vskip 6pt}\hskip 3pt\vrule width 0.6pt}\hrule height 0.6pt}\kern1pt}

\let\a=\alpha \let\b=\beta \let\g=\gamma \let\d=\delta 
  \let\q=\theta

\let\w=\omega \let\G=\Gamma   \let\L=\Lambda

\def\nn{\nonumber} \def\bd{\begin{document}} \def\ed{\end{document}}
\def\ds{\documentstyle} \let\fr=\frac \let\bl=\bigl \let\br=\bigr
\let\Br=\Bigr \let\Bl=\Bigl
\let\bm=\bibitem
\let\na=\nabla
\let\pa=\partial \let\ov=\overline
\newcommand{\be}{\begin{equation}}
\newcommand{\ee}{\end{equation}}
\def\ft#1#2{{\textstyle{{\scriptstyle #1}\over {\scriptstyle #2}}}}
\def\fft#1#2{{#1 \over #2}}
\def\vp{\varphi}
\def\sst#1{{\scriptscriptstyle #1}}
\def\oneone{\rlap 1\mkern4mu{\rm l}}
\def\td{\tilde}
\def\wtd{\widetilde}
\def\ie{\rm i.e.\ }
\def\dalemb#1#2{{\vbox{\hrule height .#2pt
        \hbox{\vrule width.#2pt height#1pt \kern#1pt
                \vrule width.#2pt}
        \hrule height.#2pt}}}
\def\square{\mathord{\dalemb{6.8}{7}\hbox{\hskip1pt}}}
\def\wtd{\widetilde}
\def\R{\rlap{\rm I}\mkern3mu{\rm R}}
\def\im{{\rm i}}
\def\tilg{\tilde{g}}
\def\tilF{\tilde{F}}
\def\tilA{\tilde{A}}
\def\varf{\varphi}
\def\tilf{\tilde{\phi}}
\def\tilh{\tilde{h}}
\def\rme{{\rm e}}
\def\te{\tilde{e}}
\def\tG{\tilde{G}}
\def\ep{\varepsilon}
\newcommand{\1}{{(1)}}
\newcommand{\2}{{(2)}}
\newcommand{\3}{{(3)}}
\newcommand{\4}{{(4)}}
\newcommand{\5}{{(5)}}
\newcommand{\xx}{{\underline{x}}}
\newcommand{\hD}{\hat{D}}
\newcommand{\hC}{\hat{C}}
\newcommand{\hg}{\hat{g}}
\newcommand{\hj}{\hat{j}}
\newcommand{\hB}{\hat{B}}
\newcommand{\hcB}{\hat{{\cal B}}}
\newcommand{\hA}{\hat{A}}
\newcommand{\hf}{\hat{\phi}}
\newcommand{\hvf}{\hat{\varphi}}
\newcommand{\hchi}{\hat{\chi}}
\newcommand{\umu}{{\underline{\mu}}}
\newcommand{\unu}{{\underline{\nu}}}
\newcommand{\urho}{{\underline{\rho}}}
\newcommand{\usigma}{{\underline{\sigma}}}
\newcommand{\ur}{{\underline{r}}}
\newcommand{\ua}{{\underline{a}}}
\newcommand{\ub}{{\underline{b}}}
\newcommand{\uc}{{\underline{c}}}
\newcommand{\ud}{{\underline{d}}}
\newcommand{\ualpha}{{\underline{\alpha}}}
\newcommand{\ubeta}{{\underline{\beta}}}
\newcommand{\ugamma}{{\underline{\gamma}}}
\newcommand{\udelta}{{\underline{\delta}}}
\begin{document}


\title{On the Stability of Anti-de Sitter Spacetime}

\author{Ali Nayeri$^{1,2}$}\author{Tuan
Tran$^{1}$} \affiliation{$^1$Institute for Fundamental Theory,
Department of Physics, University of Florida,
Gainesville, Florida 32611,\\
$^2$~~Center for Theoretical Physics,
Laboratory for Nuclear Science
and Department of Physics,\\
Massachusetts Institute of Technology, Cambridge, Massachusetts
02139}
\date{\today}

\begin{abstract}
We present a detailed analysis for the classical stability of a four
dimensional Anti-de Sitter spacetime (AdS$_4$) by decomposing the
first-order perturbations of a spherical symmetric gravitational
field into the so called tensor harmonics which transform as
irreducible representative of the rotation group (Regge-Wheeler
decomposition). It is shown that there is no nontrivial stationary
perturbation  for the angular momentum $l < 2$.  The stability
analysis forces the frequency of the gravitational modes to be
constrained in a way that the frequency of scalar modes are
constrained.
\end{abstract}

\pacs{02.40.-k, 04.20.-q, 04.25.-g}

\maketitle

\section{Introduction}
A {\it priori} there is no trivial reason to believe that a
non-hyperbolic spacetime like the Anti-de Sitter (AdS) is a stable
configuration. Stability of spacetimes with negative cosmological
constant against the quantum scalar field
\cite{AIS:1978,Burge_Lut:1985} and in the context of supergravity
\cite{Abbott_Deser:1982,Br_Freed:1982} has already been studied in
great details.  The reincarnation of the AdS spacetime in the form
of the AdS/CFT correspondence
\cite{Maldacena:1998,Gub_Kleb_Poly:1998,Witten:1998} gives a good
excuse to study the stability of the AdS spacetime in much more
details. In the present work we study thoroughly the classical
stability of the AdS$_4$ against gravitational perturbation modulo
to the ambiguity of defining a gauge invariant boundary conditions
at the timelike spatial infinity of the AdS. To this aim we use the
Regge-Wheeler metric decomposition and gauge \cite{Regge:1957} which
used initially to determine the stability of the Schwarzschild-type
black holes.  In this approach the perturbation on a spherically
background is decomposed into its normal vibrational modes using
tensor spherical harmonics. The perturbations superposed on the
AdS$_4$ background metric are the same as those given by Regge and
Wheeler \cite{Regge:1957}, consisting of odd and even parity
categories, and with the exponential time dependence, i.e.,
$\exp{(-\im\w t)}$. Thus perturbation with positive imaginary
frequencies are responsible for instability.  We show that the
frequency of the modes are discrete and real positive.  There is no
non-radiative mode  and therefore the perturbed pure AdS spacetime
is full of gravitational radiation.

The organization of our paper is as follows: In Sec. II  we review
perturbation around curved background.  In Sec. III geometrical and
group theoretical features of the four dimensional Anti-de sitter
(AdS$_4$) are studied. We parameterize AdS in the static spherical
symmetric coordinates which cover the whole spacetime. We then
perturbed this metric in accordance to Regge-Wheeler decomposition
of the metric. In Sec. IV the tensor spherical harmonics are studied
in great details. General form of the perturbation metric in
Regge-Wheeler decomposition is derived for both the odd (magnetic)
and the even (electric) modes. After gauge fixing we derive the
perturbed Einstein equations and from there the governing equations
of motion for the monopole ($l = 0$), the electric/magnetic dipole
($l = 1$) and the radiative $(l \geq 2)$ modes are studies in
details. We will argue why no non-radiative mode exists in the pure
AdS. The Schr\"{o}dinger-type equation of motion is derived for both
magnetic and electric radiative modes.  The magnetic and electric
effective potentials in the absence of any gravitational scatterer
(mass) are identical which is mainly due to the symmetry of the
AdS$_4$ group and its invariant Casimir operator. Section V is
devoted to discussion of stability of radiative modes. We find the
exact solution for the radiative modes. In order to have a
well-behaved solution in entire range of radial coordinate, the
frequency of the modes appears in discrete tones very much similar
to the frequency of the massless minimally coupled quantum scalar
field in the AdS$_4$ \cite{Burge_Lut:1985}. Finally in Sec. VI we
discuss our results and future directions.

\section{Small Fluctuations around the Curved Background}

Let us consider a general $D$-dimensional action with a cosmological
constant $\L_D$
\bea {\cal L}_D = \sqrt{{^{(D)}g}}\left[^{(D)}{R} + 2 \Lambda_{D}
\right]\;, \label{d-action} \eea
The field equations then take the following simple form
\bea ^{(D)}{R}_{MN} = -  \L_D{g}_{MN}\,. \label{general_eq} \eea

A general perturbation around this background may be written as
\bea
ds^2 = \hat{g}_{MN} dx^M dx^N = (g_{MN} + h_{MN}) dx^M dx^N\,,
\eea
where $h_{MN}$ is a small perturbation from the background $g_{MN}$.
Let us denote ``hat'' quantities to be those associated with
$\hat{g}_{MN}$ metric and quantities without ``hat'' to  those
associated with $g_{MN}$. Christoffel symbols and Ricci tensor are
defined as follows
\bea
\hat{\G}^M_{NP} & = & \fr12\,\hat{g}^{MQ}(\pa_N\hat{g}_{QP} +
\pa_P\hat{g}_{NQ} -
\pa_P\hat{g}_{NP})\,,\nn \\
\hat{R}_{MN} & = & \pa_P\hat{\G}^P_{MN} - \pa_M\hat{\G}^P_{NP} +
\hat{\G}^P_{MN}\hat{\G}^Q_{PQ} - \hat{\G}^P_{QN}\hat{\G}^Q_{PM}\,,
\eea
where
\bea \hat{g} &=& {\rm det}(\hat{g}_{MN}) = {\rm det}(g_{MN}) +
O(h^{\geq 2}) = g\,. \eea

In terms of $g_{MN}$ and up to the first order in $h_{MN}$ one has
\bea
\hat{\G}^M_{NP} = \G^M_{NP} - \fr12\,h^{MQ}(\pa_Ng_{QP} +
\pa_Pg_{NQ} - \pa_Qg_{NP}) + \fr12\,g^{MQ}(\pa_Nh_{QP} +
\pa_Ph_{NQ} - \pa_Qh_{NP})
\label{hatg-h}
\eea
Using the fact that
\bea \nabla_\a f_{\b\g} &\equiv& \pa_\a f_{\b\g} -
\G^\d_{\a\b}f_{\d\g}-\G^\d_{\a\g}f_{\d\b}\,, \eea
and
 \bea \nabla_\a f^{\b\g} &\equiv& \pa_\a f_{\b\g} +
\G^\b_{\a\d}f^{\d\g}+\G^\g_{\a\d}f_{\d\b}\,, \eea
with
\bea \G^\a_{\b\g} &\equiv& \fr12\,g^{\a\d}(\pa_\b g_{\d\g} + \pa_\g
g_{\b\d}-\pa_\d g{\b\g})\,,
\eea
One can rewrite (\ref{hatg-h}) in a compact form
\bea \hat{\G}^M_{NP} = \G^M_{NP} + \fr12\,g^{MQ}(\nabla_Nh_{QP} +
\nabla_Ph_{NQ} - \nabla_Qh_{NP}) = \G^M_{NP} + \d\G^M_{NP}\,.
\label{hatg-h1} \eea
Using the above relation and keeping only the first order in
$h_{MN}$ one finds
\bea \hat{R}_{MN} = R_{MN} + \nabla_P(\d\G^P_{MN}) -
\nabla_N(\d\G^P_{PM})\,, \eea
or
\bea \hat{R}_{MN} = R_{MN} + \fr12\,g^{PQ}(\nabla_P\nabla_Mh_{QN}+
\nabla_P\nabla_Nh_{MQ}-\nabla_M\nabla_Nh_{PQ}-\nabla_P\nabla_Qh_{MN})\,.
\label{ricci-fluc} \eea
Thus
\beq \delta R_{MN} = \frac{1}{2} \left[\Box_T h_{MN} -
\nabla^A\nabla_M h_{AN} - \nabla^A \nabla_N h_{MA} + \nabla_M
\nabla_N h^A\,_A\right] = - \L_D h_{MN} \, , \label{pert-ein}\eeq
where $\Box_T = g^{MN}\nabla_M\nabla_N$ is the covariant form of the
d'Alembert's operator, in which \bea \nabla_A \nabla_B h_{MN}& = &
\left(\del_A\del_B - \Gamma^C_{AB}\del_C\right) h_{MN} +
\left(\Gamma^C_{AM}\Gamma^E_{BN} + \Gamma^C_{AN}\Gamma^E_{BM}\right)
h_{CE} \nonumber \\
&  & - \left(\Gamma^C_{BN}\del_A + \Gamma^C_{AN}\del_B + \del_A\Gamma^C_{BN}
-\Gamma^E_{AB}\Gamma^C_{EN} - \Gamma^E_{AN}\Gamma^C_{EB}\right) h_{MC}
\nonumber \\
&  & - \left(\Gamma^C_{BM}\del_A + \Gamma^C_{AM}\del_B + \del_A \Gamma^C_{BM}
 - \Gamma^E_{AB} \Gamma^C_{EM} - \Gamma^E_{AM}\Gamma^C_{EB}\right) h_{CN} \, .
\eea

Now having (\ref{pert-ein}) in our disposal we can calculate the
perturbation modes around an AdS$_4$.  Before getting to this,
however, we first review AdS$_4$ and its geometrical and group
theoretical properties.

\section{Four dimensional Anti-\lowercase{d}e
 Sitter spacetime (A\lowercase{d}S$_4$)}
AdS$_4$ can be visualized as the four dimensional hyperboloid
\bea -U^2 - V^2 + X^2 + y^2 + Z^2 = - \ell^2 \,,\label{ads} \eea
embedded in five dimensional flat Minkowski spacetime
\bea ds^2 = -dU^2 -dV^2 + dX^2 + dY^2 + dZ^2 = \eta_{ab}\,
X^a\,X^b\,,\label{AdS_embed} \eea
where $\ell$ is the radius of the AdS$_4$ related to the Ricci
scalar curvature by $\ell^{-2} =  \left(^{(4)}R/12\right) = -
(\L_4/3)$ and $\eta_{ab} = \mbox{diag}(-,-,+,+,+)$.

To parameterize this chart we process by choosing the following
coordinates
\bea X & = & r \sin\theta \cos\varphi \,,\nonumber \\
     Y & = & r \sin\theta \sin\varphi \,,\label{X_Y_Z}\\
     Z & = & r \cos\theta \,, \nonumber \eea
where $r$,$\theta$ and $\varphi$ have their usual meaning in
spherical coordinate system. From (\ref{ads}) and (\ref{X_Y_Z}) one
finds
$$ U^2 + V^2 = \ell^2 (1 + r^2/\ell^2) \,. $$
This suggests that we can parameterize $U$ and $V$ like
\bea U & = & \ell \sqrt{1 + r^2/\ell^2} \cos{\tau} \,, \label{U}\\
     V & = & \ell \sqrt{1 + r^2/\ell^2} \sin{\tau}
     \,.\label{V}
\eea
Plugging (\ref{X_Y_Z}), (\ref{U}) and (\ref{V}) into
(\ref{AdS_embed}) gives
\bea ds^2 & = & -\ell^2 (1 + r^2/\ell^2) d\tau^2 - \frac{r^2 d
r^2}{\ell^2 (1+ r^2/\ell^2)} + dr^2 + r^2 (d\theta^2 + \sin^2\theta
d
\varphi^2) \,, \nonumber \\
& = &  -\ell^2 (1 + r^2/\ell^2) d\tau^2 + \frac{d r^2}{(1+
r^2/\ell^2)}+ r^2 (d\theta^2 + \sin^2\theta d \varphi^2)\,.\eea
By rescaling $\tau \rightarrow (t/\ell)$ we finally find a
spherically symmetric Schwarzschild type metric for AdS$_4$
\bea ds^2_{AdS_4} = -(1 + r^2/\ell^2) d t^2 + \frac{d r^2}{(1+
r^2/\ell^2)}+ r^2 (d\theta^2 + \sin^2\theta d
\varphi^2)\,.\label{ads_shcwar}
\eea
This parametrization covers the entire spacetime. As can be seen
from (\ref{X_Y_Z}), (\ref{U}) and (\ref{V}) the topology of AdS$_4$
is $S^1$ (time) $\times$ $R^3$ (space) which signals that in AdS
there are closed timelike lines.  In order to get rid of this
sickness one has to unwrap the $S^1$ and work, instead, in covering
spacetime (CAdS$_4$) with topology $R^4$ with no closed timelike
lines. In this parametrization of AdS, the spatial infinity is
timelike and thus information can be lost to, or gained from it.
Unfortunately any change in time coordinate to get rid of this
problem will cost us to lose globally defined coordinates.

Now for the sake of generality we write the metric in the generic
form of spherical symmetric metric, i.e.,
\bea ds_4^2 = - \rme^{A(r)} dt^2 + \rme^{B(r)} dr^2 + r^2 d\theta^2
+ r^2 \sin^2\theta d\varphi^2 \eea
where $A, B$ are functions of $r$ only. The non-vanishing components
of Christoffel symbols are
\bea \G_{tr}^t &=& \fr{A^\prime}{2}\;,\;\;
\G_{tt}^r=\rme^{A-B}\fr{A^\prime}{2}\;,\;\;
\G_{rr}^r=\fr{B^\prime}{2}\;,\;\;\G_{\theta\theta}^r=-r\,\rme^{-B}\;,\;\;
\G_{\varphi\varphi}^r=-r\sin^2\theta\,\rme^{-B}\nonumber \\
\G_{r\theta}^\theta&=&\fr1{r}\;,\;\;\G_{\varphi\varphi}^\theta=-\sin\theta\cos\theta\;,\;\;
\G_{r\varphi}^\varphi=\fr1{r}\;,\;\;\G_{\theta\varphi}^\varphi=\cot\theta\,,
\eea
where prime means derivative with respect to $r$.

The group $SO(2,3)$ is the symmetry group of the $4$-dimensional
Anti-de Sitter spacetime which plays the role of the $4$-dimensional
Poincar\'{e} group.  The time translations are the $U(1)$ subgroup
of rotations in the $(U,V)$-plane.  The $10$ generators denoted by
$J_{ab}$ act on the embedding coordinates and are defined by
\bea J_{ab} = X_a \frac{\partial}{\partial X^b} - X_b
\frac{\partial}{\partial X^a}\,,
\eea
where $X_a = \eta_{ab}\,X^b$.  Thus the generators $J_{ab}$ satisfy
\bea [J_{ab},J_{cd}] = \eta_{ad}\,J_{bc} + \eta_{bc}\,J_{ad} -
\eta_{ac}\,J_{bd} - \eta_{bd}\,J_{ac}\,.
\eea
Then $J_{\mu\nu}$ generate ordinary rotations around any points in
the four space spanned by $U, X, Y$ and $Z$. The translational
operators are defined by
\bea
P_\mu = \ell^{-1} L_{\mu V} \,,
\eea
for $\mu = U, X, Y$ and $Z$.  The commutation relations between
$L_{\mu\nu}$ and $P_\mu$ are
\bea [L_{\mu\nu},P_\rho]& = & \eta_{\mu\rho}\,P_\nu -
\eta_{\nu\rho}\,P_\mu \,,\nonumber  \\
\ [P_\mu,P_\nu] & = & \frac{1}{\ell^2}L_{\mu\nu}\,. \eea
These results hold for any spin and reduce to the commutation
relations for the Poincar\'{e} group when $\ell \longrightarrow
\infty$ . The Casimir bilinear invariant operator $C_2$ can be
written as
\bea C_2 = \frac{1}{2} L_{ab}\,L^{ab} =
\frac{1}{\ell}\frac{\partial}{\partial
X^a}X_a\,X_b\frac{\partial}{\partial X^b} - \eta^{ab}
\frac{\partial}{\partial X^a}\frac{\partial}{\partial X^b} + s(s +
1) \,, \eea
where $s$ is the spin.  Note that the first two term in $C_2$ is the
same as the invariant Laplace-Beltrami operator.  Thus
$\Box_{AdS_4}$ is an invariant with respect to $SO(2,3)$ and a
function of $C_2$ \cite{Fronsdal:1974,Fronsdal:1975}.  For the
background (\ref{ads_shcwar}), $C_2$ takes the following form for
the gravitons
\bea C_2 = \left[1 - \frac{1}{\ell^2}(r^2 - t^2)\right]\left[-
\Box_s + \frac{1}{\ell^2}\left(t\frac{\partial}{\partial t} +
r\frac{\partial}{\partial r} + 1\right)^2\right] + 6 \,, \eea
where $\Box_s$ is the Laplace-Beltrami operator in spherical
coordinates.

\section{Fluctuations around A\lowercase{d}S$_4$ background}
Perturbations can be expressed in the form of four factors, each
of which is a function of the coordinates, $t$, $r$, $\theta$, and
$\varphi$; this separation is achieved by the use of generalized
tensor spherical harmonics. The ten components of metric
perturbations can be divided into two categories, called
perturbations of odd (magnetic) and even (electric) parity, which
are not mixed by tensorial operators that respect spherical
symmetry. Under a rotation of the frame around the origin, the
$\fr{1}{2}D(D + 1)$ components of the perturbing metric transform
like 3 scalars: $(h_{00}, h_{01}, h_{11})$, and 2 $(D-2)$-vectors:
$(h_{02}, h_{03},...,h_{0D-1}; h_{12}, h_{13},...,h_{1D-1})$, and
a $\fr{1}{2}(D - 1)(D - 2)$ component second rank tensor. For
instance, when $D = 4$ we have the following blocks,
\bea h_{\mu\nu} = \left[
\begin{array}{cc}
\stackrel{\mathrm{Scalars}}{\fbox{$\displaystyle
\begin{array}{cc}
h_{00} & h_{01} \\
h_{10} & h_{11}
\end{array}
$}} &
\stackrel{\mathrm{Vectors}}{\fbox{$\displaystyle
\begin{array}{cc}
h_{02} & h_{03} \\
h_{12} & h_{13}
\end{array}
$}}\\
\stackrel{\mathrm{Vectors}}{\fbox{$\displaystyle
\begin{array}{cc}
h_{20} & h_{21} \\
h_{30} & h_{31}
\end{array}
$}} &
\stackrel{\mathrm{Tensors}}{\fbox{$\displaystyle
\begin{array}{cc}
h_{22} & h_{23} \\
h_{32} & h_{33}
\end{array}
$}}
\end{array}
\right]\,, \eea
whereas for $D = 5$ we get,
\bea h_{AB} = \left[
\begin{array}{cc}
\stackrel{\mathrm{Scalars}}{\fbox{$\displaystyle
\begin{array}{cc}
h_{00} & h_{01} \\
h_{10} & h_{11}
\end{array}
$}} & \stackrel{\mathrm{Vectors}}{\fbox{$\displaystyle
\begin{array}{ccc}
h_{02} & h_{03} & h_{04}\\
h_{12} & h_{13} & h_{14}\\
\end{array}
$}}\\
\stackrel{\mathrm{Vectors}}{\fbox{$\displaystyle
\begin{array}{cc}
h_{20} & h_{21}\\
h_{30} & h_{31}\\
h_{40} & h_{41}
\end{array}
$}} & \stackrel{\mathrm{Tensors}}{\fbox{$\displaystyle
\begin{array}{ccc}
h_{22} & h_{23} & h_{24} \\
h_{32} & h_{33} & h_{34} \\
h_{42} & h_{43} & h_{44}
\end{array}
$}}
\end{array}
\right]\,. \eea

Associated with any perturbation modes, there is the angular
momentum $l$ and its projection on the $z$ axis $m$. The general
perturbation is expressed by $h_{\mu\nu} = \sum_{l m}
[h_{\mu\nu}^{(odd) lm} + h_{\mu\nu}^{(even) lm}]$, where
$h_{\mu\nu}^{(odd) lm}$ and $h_{\mu\nu}^{(even) lm}$ behave
differently under parity change: ${\cal
P}[h_{\mu\nu}^{lm}(t,r,\theta,\varphi)] \rightarrow
\tilde{h}_{\mu\nu}^{lm}(t,r,\pi - \theta, \pi + \varphi)$. In
practice, a tensor harmonic is {\it odd} or {\it axial} if ${\cal
P}(h_{\mu\nu}) = \tilde{h}_{\mu\nu}= (-1)^{l + 1}h_{\mu\nu}$ and is
{\it even} or {\it polar} if ${\cal P}(h_{\mu\nu}) =
\tilde{h}_{\mu\nu}= (-1)^{l}h_{\mu\nu}$. In four dimensions, as we
mentioned above, we have three scalars under rotation,
\be s_{\mu\nu}(t,r,\theta,\phi) = \sum_{l m}
\alpha_{\mu\nu}^{lm}(t,r) Y^m_l(\theta, \varphi) \,, \ee
where $Y^m_l(\theta,\varphi)$s are well known scalar spherical
harmonics and are given by
\bea Y^m_l(\theta,\varphi) = \left[\frac{(2 l + 1)(l -m)}{4\pi(l +
m)!}\right]^{(1/2)} P^m_l(\cos\theta) \rme^{i m\varphi}\,,\eea
and
\bea \alpha_{\mu\nu}^{lm}(t,r) = e^A
H_0^{lm}(t,r)\delta^t\;_\mu\delta^t\;_\nu +
H_1^{lm}(t,r)(\delta^t\;_\mu\delta^r\;_\nu + \delta^r\;_\mu
\delta^t\;_\nu) + e^B H_2^{lm}(t,r)\delta^r\;_\mu\delta^r\;_\nu\,.
\eea
For vector spherical harmonics \cite{Thorne:1980} we have two
distinct types with opposite parities, i.e,
\bea {Y^{ml}_{a} (\theta,\varphi) = \left\{\begin{array}{ll} &
[l(l + 1)]^{(-1/2)} \frac{\partial Y^m_{l}}{\partial x^a}\,;
\,\,\, \mbox{with
${\cal P} =(-1)^l$} \\
& [l(l + 1)]^{(-1/2)} \epsilon_a\;^b\;\frac{\partial
Y^m_{l}}{\partial x^b}\,, \,\,\, \mbox{with ${\cal P} = (-1)^{l +
1}$}
\end{array}\right.} \eea
where $\epsilon_{ab}$ is totally antisymmetric tensor which is
covariantly constant on $S^2$, i.e, $\epsilon_{bc\;;a} = 0$ and is
defined by ,
\bea \epsilon_{\theta\theta} & = & \epsilon_{\varphi\varphi} = 0
\nonumber \\
\epsilon_{\theta\varphi} & = & -\epsilon_{\varphi\theta} =
\sin\theta \;.\eea
Here the lower case Latin letters $a$, $b$, and $c$ run over the
values $\theta$ and $\varphi$.

The vector part of the metric then is given by
\bea v_{\mu\nu}(t, r, \theta, \varphi) = \sum_{lm}
\left[\alpha^{(lm)\;a}_{\mu\nu}(t,r)\;Y^{lm
(odd)}_a(\theta,\varphi) + \beta^{(lm)\;a}_{\mu\nu}(t,r) \; Y^{lm
(even)}_a (\theta,\varphi)\right]\,,\eea
where
\bea \alpha^{(lm)\;a}_{\mu\nu}(t,r) & = & \sqrt{l(l + 1)}
\left[h_0^{lm(odd)}(t,r)\delta^t\;_\mu +
h_1^{lm(odd)}(t,r)\delta^r\;_\mu\right]\delta^a\;_\nu \nonumber
\\ & & + \sqrt{l(l + 1)}\,\delta^a\;_\mu\left[h_0^{lm (odd)}(t,r) \delta^t\;_\nu +
h_1^{lm (odd)}(t,r) \delta^r\;_\nu\right]\,, \eea
and
\bea \beta^{(lm)\;a}_{\mu\nu}(t,r) & = & \sqrt{l(l + 1)}
\left[h_0^{lm(even)}(t,r)\delta^t\;_\mu +
h_1^{lm(even)}(t,r)\delta^r\;_\mu\right]\delta^a\;_\nu \nonumber
\\
& & + \sqrt{l(l + 1)}\,\delta^a\;_\mu\left[h_0^{lm (even)}(t,r)
\delta^t\;_\nu + h_1^{lm (even)}(t,r) \delta^r\;_\nu\right]\,,\eea
with $a$ runs over $\theta$ and $\varphi$.

For a rank-2 symmetric tensor, there are three fundamental types
of tensor of angular momentum $l$\cite{Regge:1957},
\bea {Y^{lm}_{ab} = \left\{\begin{array}{lll} \psi^{lm}_{ab} & = &
(1/\sqrt{l(l + 1)[l(l + 1) - 1]})\,\, Y^m\;_{l\;|ab}\;;
\,\,\, \mbox{with ${\cal P} = (-1)^l$}\\
\phi^{lm}_{ab} & = & \frac{1}{\sqrt{2}} \gamma_{ab}Y^m_l \;;\,\,\,\,\,\,\,\,\, \mbox{with ${\cal P} = (-1)^l$} \\
\chi^{lm}_{ab} & = & \sqrt{\frac{(l - 2)!}{2 (l + 2)!
}}\left[\epsilon_a\;^c\; Y^m\;_{l\;|cb} + \epsilon_b\;^c\;
Y^m\;_{l\;|ca}\right]\;;\,\,\, \mbox{with ${\cal P} = (-1)^{l+1}$
and zero trace}\,,
\end{array}\right.}\eea
where $\gamma_{ab} = g_{ab}/r^2$ is the metric tensor and the $|$
denotes covariant derivative on $S^2$.  The tensor part of the
metric then takes the following form
\bea t_{\mu\nu}(t, r, \theta, \varphi) = \sum_{lm}\left[
\alpha^{(lm)\;ab}_{\mu\nu}(t,r) \chi^{lm
(odd)}_{ab}(\theta,\varphi) + \beta^{(lm)\;ab}_{\mu\nu}(t,r)
\phi^{lm (even)}_{ab}(\theta,\varphi) +
\beta'^{(lm)\;ab}_{\mu\nu}(t,r) \psi^{lm
(even)}_{ab}(\theta,\varphi)\right]\,,\eea
where
\bea \alpha^{(lm)\;ab}_{\mu\nu}(t,r) = \sqrt{\frac{(l + 2)! }{2 (l
- 2)!}} h_2^{lm}(t,r)\,\delta^a\;_\mu \delta^b\;_\nu \,,\eea
and
\bea \beta^{(lm)\;ab}_{\mu\nu}(t,r) & = & r^2 \sqrt{2}
K^{lm}(t,r)\;
\delta_\mu\;^a \; \delta_\nu\;^b\,, \\
\beta'^{(lm)\;ab}_{\mu\nu}(t,r)& = & r^2 \sqrt{l(l + 1)[l (l +1) -
1]}\; G^{lm}(t,r) \; \delta_\mu\;^a\;\delta_\nu\;^b
 \,. \eea

Now we can summarize the results we have just derived.  The most
general form of the perturbed metric for the two distinct parities
are

\noindent {\bf odd parity:}
\bea h^{(odd)}_{\mu\nu} = \left[\begin{array}{cccc}
  0 & 0 & -h_0(r,t)\left(\fr{\partial}{\sin\theta\partial \varphi}\right)Y^m_l &  h_0(r,t)\sin\theta\left(\fr{\partial}{\partial \theta}\right)Y^m_l\\
  0 & 0 & -h_1(r,t)\left(\fr{\partial}{\sin\theta\partial \varphi}\right)Y^m_l &  h_1(r,t)\sin\theta\left(\fr{\partial}{\partial \theta}\right)Y^m_l \\
  h_{t\theta}& h_{r\theta}& h_2(r,t)\left[\fr{\partial^2}{\sin\theta\partial\varphi\partial\theta} - \cos\theta\fr{\partial}{\sin^2\theta\partial\varphi}\right]Y^m_l & h_{\varphi\theta} \\
  h_{t\varphi}& h_{r\varphi} & \fr{1}{2}h_2(r,t)\left[\fr{\partial^2}{\sin\theta\partial\varphi^2} + \cos\theta\fr{\partial}{\partial\theta}-
  \sin\theta\fr{\partial^2}{\partial\theta^2}
\right]Y^m_l & -
h_2(r,t)\left[\sin\theta\fr{\partial^2}{\partial\theta\partial\varphi}
- \cos\theta\fr{\partial}{\partial\varphi}\right]Y^m_l
\end{array}
\right]\;; \eea
\noindent {\bf even parity:}
\bea h^{(even)}_{\mu\nu} = \left[\begin{array}{cccc}
  e^AH_0(t,r)Y^m_l & H_1(t,r)Y^m_l & h_0(r,t)\left(\fr{\partial}{\partial\theta}\right)Y^m_l &  h_0(r,t)\left(\fr{\partial}{\partial\varphi}\right)Y^m_l\\
  H_1(t,r)Y^m_l& e^BH_2(t,r)Y^m_l & h_1(r,t)\left(\fr{\partial}{\partial\theta}\right)Y^m_l &  h_1(r,t)\left(\fr{\partial}{\partial\varphi}\right)Y^m_l\\
  h_{t\theta}& h_{r\theta}& r^2\left[K(t,r) + G(t,r)\fr{\partial^2}{\partial\theta^2}\right]Y^m_l & h_{\varphi\theta} \\
  h_{t\varphi}& h_{r\varphi} & r^2G(t,r)\left[\fr{\partial^2}{\partial\theta\partial\varphi}\right. & r^2\left[K(t,r)\sin^2\theta \right.\\
& & \left.- \cos\theta\fr{\partial}{\sin\theta\partial\varphi}
\right]Y^m_l& +\left.
G(t,r)\left(\fr{\partial^2}{\partial\varphi^2} + \sin\theta
\cos\theta\fr{\partial}{\partial\theta}\right)\right]Y^m_l
\end{array}
\right]\;. \label{even_general_metric}\eea

\subsection{Magnetic (odd) modes}
We can now simplify the perturbations by the gauge transformation
due to the coordinates transformation $\hat x^\mu = x^\mu +
\xi^\mu$ with $\xi^\mu \ll x^\mu$:
\bea \hat {h}_{\mu\nu}(x^\rho)& = & h_{\mu\nu}(x^\rho) -
\xi_{\mu;\nu} - \xi_{\nu:\mu} \nn \\
& = & h_{\mu\nu}(x^\rho) - g_{\sigma\nu}\xi^\sigma\,_{,\mu} -
g_{\mu\sigma}\xi^\sigma\,_{,\nu} - \xi^\sigma g_{\mu\nu,\sigma}\,.
\eea
We choose a gauge transformation that eliminates those terms which
contain the derivatives of the highest order with respect to
angles \footnote{One of the most popular gauge in studying tensor
perturbation in general relativity is the so-called transverse
traceless gauge, i.e., $h^{\mu\nu}\;_{;\nu} = 0 = h^{\mu}\;_{\mu}
$.  On $S^2$, however, it is well known that no second rank
symmetric tensor spherical can exist \cite{Higuchi:1987}. Using
spherical harmonics to decompose the metric in $4$ dimensions do
not permit us to use transverse traceless gauge.}:
\bea \xi^\mu_{(odd)} = A(t,r) \delta^\mu\/_a
\epsilon^{ab}\fr{\partial}{\partial x^b} Y^m_l\;\;\;\;\;
\mbox{(with $a, b = \theta,\varphi$)}\,. \label{odd_gauge} \eea
This specialization is accomplished by demanding radial function
$A(t, r)$ to annul the radial factor $h_2(t,r)$, i.e, by choosing
$A = \frac{1}{2}(h_2^{(odd)}/r^2)$. Also since the final result is
independent of particular value of $m$, we choose $m = 0$. Finally
we take the time dependence of the perturbations as $\exp{(-\im \w
t)}$, since the background is independent of time.  Note that
purely positive imaginary frequencies that make perturbations grow
exponentially with time. Thus the non-trivial {\it odd} modes are
taken to be
\bea h^{(odd)}_{t\varphi} = h_0(r)\,f(\theta)\,\rme^{-i\w
t}\;,\;\; h^{(odd)}_{r\varphi} = h_1(r)\,f(\theta)\,\rme^{-i\w
t}\, \label{odd-modes-pert-metric} \eea
where, $ f(\theta) = \sin\theta\,(d/d\theta)P_l(\theta)$.  Thus
the {\it odd} part of the metric is
\bea h^{(odd)}_{\mu\nu} = \left[
\begin{array}{cccc}
0& 0 & 0 & h_0 \\
0& 0 & 0 & h_1 \\
0 & 0 & 0 & 0 \\
h_0 & h_1 & 0 & 0
\end{array}
\right]\,\rme^{-\im\,\w t}\,f(\theta)\,. \eea
There are ten equations of motion but only three of them are
non-trivial. The non-trivial equations of motion according to
(\ref{pert-ein}) are
\bea \d R_{r\varphi} + \L_4 h_{r\varphi} =  0   &\rightarrow &
 \left\{
\left[\fr1{r}(A^\prime-B^\prime)\,\rme^{-B} + \w^2\,\rme^{-A} +
\fr2{r^2}\,\rme^{-B} -\fr{6}{\ell^2} -
\fr{l(l+1)}{r^2}\right]\,h_1 \right.\nn \\
& & \left. +\im\,\w\rme^{-A}\left(\fr{2}{r}-
\fr{d}{dr}\right)h_0\right\}\sin\theta\,
\fr{dP_l(\theta)}{d\theta}\,\rme^{-\im\w t} =
0\,,\label{eq-per-13}\\
\d R_{\theta\varphi} + \L_4 h_{\theta\varphi} =  0  &\rightarrow&
\left\{\im\,\w\,h_0\,\rme^{-A} +
 \rme^{-B}\,\left[\fr{A^\prime - B^\prime}{2} +
\fr{d}{dr}\right]\,h_1\right\} \times
\left(\cos\theta\,\fr{d}{d\theta} -
\sin\theta\,\fr{d^2}{d\theta^2}\right)P_l(\theta)\,\rme^{-\im\w
t}\,,
\label{eq-per-23}\\
\d R_{t\varphi}+ \L_4 h_{t\varphi} = 0 &\rightarrow&
\left\{\rme^{-B}\left[\fr{d^2}{dr^2}-
\fr{A^\prime+B^\prime}{2}\,\fr{d}{dr} -\fr{l(l+1)}{r^2}\,\rme^{B}
+\fr{2}{r}A^\prime -\fr{3}{\ell^2}\,\rme^{B}\right]\,h_0 \right.
\nn \\
&  & \left. + \im\,\w\,\rme^{-B}\,\left[\fr{d}{dr} + \fr2{r}-
\fr{A^\prime + B^\prime}{2}\right]\,h_1\right\}
\sin\theta\,\fr{dP_l(\theta)}{d\theta}\,\rme^{-\im\w t} = 0\, .
\label{eq-per-03} \eea

{\bf I. Special case, $l = 0$ :}

All the angular factors in Eqs.~(\ref{eq-per-13}),
(\ref{eq-per-23}), and (\ref{eq-per-03}) are identical to zero for
$l = 0$, and since $f(\theta) = 0$ it is clear that
$[h_{\mu\nu}^{l=0, m =0}]^{(odd)} = 0$.

{\bf II. Special case, $l = 1$: }

For $l = 1$, however, $\d R_{\theta\varphi}$ is trivial and $\d
R_{r\varphi}$ yields the following relation between $h_0$ and $h_1$,
\bea h_1 =
\frac{\im}{\w}r^2\frac{d}{dr}\left(\frac{h_0}{r^2}\right)\,, \eea
where the time dependence of the perturbations is retained as
$\exp{(-\im \w t)}$. Now one can show that these two related modes
can be gauged away through gauge transformation (\ref{odd_gauge}),
by choosing $A$ to be $(i/\omega)h_0$:
\be \xi_{\varphi}= \left(\fr{i}{\omega}\right) h_0 \exp{(- i
\omega t)} \sin\theta \fr{d}{d\theta}P_1(\cos\theta)\,. \ee
This gauge transformation leaves other perturbation components
unaltered. Thus $[h_{\mu\nu}^{l= 1, m}]^{(odd)} = 0$.

{\bf III. Case $l \geq 2$:}

From Eq.~(\ref{eq-per-13}) and (\ref{eq-per-23}), one can
eliminate function $h_0$ in terms of $h_1$. The result is
\bea \rme^{-B}\left[\fr{d^2}{dr^2} + \fr{A^{\prime\prime} -
  B^{\prime\prime}}{2} + \fr{3(A^\prime-B^\prime)}{2}\fr{d}{dr} +
  \fr{(A^\prime - B^\prime)^2}{2} - \fr2{r^2}\fr{d}{dr} +
  \fr2{r^2}\right]\,h_1
+ \left[ -\fr{3}{\ell^2} - \fr{l(l+1)}{r^2} +
\rme^{-A}\,\w^2\right]h_1 = 0 \label{eq-per-123} \eea
Define
\bea {\cal M} & = & \rme^{\fft12(A-B)}\,\fr{h_1}{r}\,,\nn \\
 dr^* & = & \rme^{\fft12(B-A)}\,dr\,, \label{q-def} \eea
so that
\bea \fr{d^2{\cal M}}{d{r^*}^2} &=& \fr{\rme^{\fft32(A-B)}}{r}\,
\left[h_1^{\prime\prime} + \fr{3(A^\prime-B^\prime)}{2}h_1^\prime
+ \fr{A^{\prime\prime}-B^{\prime\prime}}{2}h_1 +\fr2{r^2}\,h_1 +
\fr{(A^\prime-B^\prime)^2}{2}h_1 - \fr{2}{r}\,h_1^\prime\right]\nn\\
&&-\rme^{\fft32(A-B)}\,\fr{3(A^\prime - B^\prime)}{2r}h_1\,.
\label{q-derivative} \eea
Using (\ref{q-def}) and (\ref{q-derivative}), we rewrite
Eq.~(\ref{eq-per-123}) as follows
\bea \fr{d^2{\cal M}}{d{r^*}^2} + (\w^2  - V_{\rm
eff}^{odd})\,{\cal M} = 0\,, \label{odd_schrod}\eea
where
\bea V_{\rm eff}^{odd} = \fr{6}{\ell^2}\,\rme^{A} +
\fr{l(l+1)}{r^2}\,\rme^{A} -
\fr{3(A^\prime-B^\prime)}{2r}\rme^{A-B}\,. \eea
For the $AdS_4$, one has
\bea \rme^{A} = 1 + \fr{r^2}{\ell^2} = \rme^{-B}\,, \eea
with
\bea r = \ell \tan{\left(\fr{r^*}{\ell}\right)} \,,\eea
which means $r^* \in [0, \pi/2]$. So the effective potential can
be written as
\bea V_{\rm eff}^{odd} & = & \left(1 +
\frac{r^2}{\ell^2}\right)\left[\fr6{\ell^2} + \fr{l(l+1)}{r^2}-
\fr{6}{\ell^2} \right]\nn \\
& = & \left(1 +
\frac{r^2}{\ell^2}\right)\fr{l(l+1)}{r^2}\label{eff-pot}\,. \eea
Thus the effective potential $V_{\rm eff}^{odd}$ is real and
positive everywhere

\subsection{Even (electric) modes}
For the electric parity, in order to fix the coordinates to the
first order, we demand that the even parity functions $h_0$, $h_1$
and $G$ vanish.  This can be achieved by the following gauge
transformations:

\bea \xi^t_{even} & = & \rme^{-A} \left(\fr{1}{2} r^2 \fr{\partial
G(t, r)}{\partial t} - h_0\right)Y^M_l(\theta,\varphi)\,, \\
\xi^r_{even} & = & \rme^A \left(-\fr{1}{2} r^2 \fr{\partial
G(t, r)}{\partial t} + h_1\right)Y^M_l(\theta,\varphi)\,,\\
\xi^\theta_{even} & = & \fr{1}{2} G(t, r) \fr{\partial Y^M_l(\theta,\varphi)}{\partial \theta}\,,\\
\xi^\varphi_{even} & = & \fr{1}{2} G(t, r) \fr{1}{\sin^2\theta}
\fr{\partial Y^M_l(\theta,\varphi)}{\partial \varphi}\,. \eea

Therefore the perturbed metric for the even parity waves in the
canonical form (and by specializing to $m = 0$) is
\bea h_{\mu\nu}^{even} = \left[
\begin{array}{cccc}
H_0(r)\,\rme^{A} & H_1(r) & 0 & 0 \\
H_1(r) & H_2(r)\,\rme^{B} & 0 & 0 \\
0 & 0 & K(r)r^2 & 0 \\
0 & 0 & 0 & K(r)\,r^2\,\sin^2\theta
\end{array}
\right]\,\rme^{-\im\,\w t}\,P_L(\q)\,.
\eea
There are 7 non-trivial equations of motion for four unknowns
($H_0$, $H_1$, $H_2$, and $K$) which correspond to four diagonal
components of Ricci tensor and three off-diagonal components of
Ricci tensor ($(t,r), (t,\theta), (r,\theta)$). Explicitly, they are
\bea \delta R_{tr} + \L_4 h_{tr} &=& \left\{\im\, \w\,\left(K^\prime
- \fr12\,KA^\prime + \fr{K}{r} -
\fr1{r}\,H_2\right)\right.\nn\\
&& \left.+ \rme^{-B} \left[\fr{L(L+1)}{2r^2}\,\rme^{B}
+\fr{3}{\ell^2}\,\rme^B -\fr{A^{\prime\;2}}{4} + \fr{A^\prime
B^\prime}{4} -\fr{A^\prime}{r} - \fr{A^{\prime\prime}}{2}
\right]\,H_1\right\}\,P_L(\theta)\,\rme^{-\im\w t} = 0\\[0.6cm]
\delta R_{t\theta} + \L_4 h_{t\theta} &=& \left\{ \im \,\w\,(K +
H_2) + \fr{A^\prime - B^\prime}{2}\,H_1\,\rme^{-B} +
H_1^\prime\,\rme^{-B}\right\}\,
\fr{dP_L(\q)}{d\q}\,\rme^{-\im\w t} = 0 \label{R_0_2_even}\\[0.6cm]
\delta R_{r\theta} + \L_4 h_{r\theta} &=& \left\{\im
\,\w\,H_1\,\rme^{-A} + H_0^\prime-K^\prime +
\left(\fr{A^\prime}{2}-\fr1r\right)H_0\right.\nn\\
&&\left. +\left(\fr{A^\prime}{2}+\fr1r\right)H_2
\right\}\,\fr{dP_L(\q)}{d\q}\,\rme^{-\im\w t} = 0\\[0.6cm]
\delta R_{tt} + \L_4 h_{tt}&=& \left\{\fr12\,\w^2 (H_2+2K) + \im
\w\,\rme^{-B}\,\left(\fr{B^\prime}{2} - \fr{2}{r} -
\fr{d}{dr}\right)H_1 -
\fr12\,\rme^{A-B}\,H_0^{\prime\prime}\right.\nn\\
&& + \rme^{A-B}\,\left(\fr{B^\prime}{4} - \fr{A^\prime}{2} - \fr1r\right)H_0^\prime
-\fr14\,\rme^{A-B}\,A^\prime H_2^\prime + \fr12\,\rme^{A-B}\,A^\prime K^\prime\nn\\
&&+ \rme^{A-B}\left[\fr{L(L+1)}{2r^2}\,\rme^{B} +
\rme^{B}\,\fr{3}{\ell^2} -\fr{A^\prime}{r} - \fr{A^{\prime\;2}}{4}
+ \fr{A^\prime B^\prime}{4} -
\fr{A^{\prime\prime}}{2}\right]H_0\nn\\
&&\left.+\rme^{A-B}\,\left(\fr{A^\prime B^\prime}{4} -
\fr{A^{\prime\prime}}{2} - \fr{A^{\prime\;2}}{4} -
\fr{A^\prime}{r}\right)\,H_2\right\}\,
P_L(\theta)\,\rme^{-\im\w t} = 0\,,\\[0.6cm]
\delta R_{rr} + \L_4 h_{rr} &=& \left\{
\im\,\w\,\rme^{-A}\left(H^\prime_1 - \fr{B^\prime}{2}H_1\right)
- \fr12\,\rme^{B-A}\,\w^2\,H_2\right.\nn\\
&& + \fr{H_0^{\prime\prime}}{2} -
K^{\prime\prime} + \left(\fr{A^\prime}{2} -
\fr{B^\prime}{4}\right)\,H_0^\prime + \left(\fr{A^\prime}{4} +
\fr1{r}\right)\,H_2^\prime \nn\\
&&\left.+\left(\fr{B^\prime}{2} - \fr2{r}\right)\,K^\prime +
\rme^{B}\left[\fr{L(L+1)}{2r^2} + \fr{3}{\ell^2}\right]H_2
\right\}\,P_L(\theta)\,\rme^{-\im\w t}
 = 0 \,, \\[0.6cm]
\delta R_{\theta\theta} + \L_4 h_{\theta\theta} & =& \delta
R_{\varphi\varphi}+ \L_4 h_{\varphi\varphi} =  \nn \\
& = & \left\{-\fr{\w^2r^2}{2}\rme^{-A}K + \im\w\rme^{-A-B}\,r\,H_1 -
\fr12\,\rme^{-B}r^2\,K^{\prime\prime} +
\fr12\,r\rme^{-B}\,H_0^\prime
+ \fr12\,r\rme^{-B}\,H_2^\prime\right.\nn\\
&& + \rme^{-B}\,\left(\fr{r^2B^\prime}{4} - \fr{r^2A^\prime}{4}
-2r\right)\,K^\prime + \rme^{-B}\,\left(\fr{rA^\prime}{2} -
\fr{rB^\prime}{2} + 1\right)\,H_2\nn\\
&&\left.+\rme^{-B}\,\left[ \fr{L(L+1)}{2}\,\rme^{B} +
\fr{3}{\ell^2}\,\rme^B
-1+\fr{r(B^\prime-A^\prime)}{2}\right]K\right\}\,
P_L(\q)\,\rme^{-\im\w t}\nn\\
&& + \fr12\,(H_0-H_2)\fr{d^2P_L(\q)}{d\q^2}\,\rme^{-\im\w t} = 0 \,.
\eea

{\bf I. Special case, $l = 0$ :}

Making a gauge transformation $\xi^t_{even} = E_0(t,r) Y^0_0$ and
$\xi^r_{even} = E_1(t,r) Y^0_0$ we can choose $E_0(r,t)$ and
$E_1(r,t)$ such that $H_1(r, t) = 0 = K(r, t)$, i.e.,
\bea \xi^t & = & \int\;dr'
\rme^{-A(r)}\left(\frac{1}{2}r\rme^{-A(r')}
\frac{\partial K(t,r')}{\partial t} - H_1(t,r')\right) + C(t)\,, \\
\xi^r & = &\frac{1}{2} r K(t,r)\,.\eea
Since $Y^0_0$ is a number, the only trivial magnetic equations for
$l = 0$ are $\delta R_{tr}$, $\delta R_{tt}$, $\delta R_{rr}$.
Equation $\delta R_{\theta\theta} = 0$ is satisfied identically by a
solution of other three equations. Equation $\delta R_{tr} = 0$
assures us that $H_2(t, r) = H_2(r)$ and $\delta R_{tt} + \delta
R_{rr} = 0$ gives $\partial H_0(t, r)/\partial r = d H_2(r)/d r$,
i.e.,
\bea H_0(r,t) = H_2(r) + c(t)\,, \eea
where $c(t)$ is some arbitrary function of time and can be removed
by a suitable gauge transformation of the form $\xi^t =
\frac{1}{2} Y^0_0 \int c(t') dt'$.  Thus $H_0(r) = H_2(r)$.

Note that for the $AdS_4$ background we have
\[ A^{\prime\prime} + A^{\prime\;2} + \fr2{r}A^\prime = \fr6{\ell^2}\rme^{-A}\,.\]
Equation $\delta R_{\theta\theta} = 0$ now gives
\bea H_0(r) = H_2(r) = \frac{c}{r(1 + r^2/\ell^2)}
\label{even_elec_pot}\,,\eea
where $c$ is a constant which has to be determined according to
the boundary conditions.  Though Eq. (\ref{even_elec_pot}) well-
behaves at infinity it blows up at the origin, so the suitable
choice for the $c$ is to set it to zero which means,
\bea H_0 = H_2 = 0 \longrightarrow [h_{\mu\nu}^{l = 0, m =
0}]^{(even)} = 0 \,.\eea
This proves that the monopole perturbation on $AdS_4$ is zero as one
may expect due to the absence of massive perturbation.  So the
conclusion is, for $AdS_4$ background, $h_{\mu\nu}^{00} = 0$.

{\bf II. Special case: $l = 1$:}

For $l = 1$ one can readily see from (\ref{even_general_metric})
that there are only six independent radial functions in the general
form, i.e., the triviality of the angular part in $h_{\theta\varphi}
= h_{\varphi\theta}$ leaves $G^{lm}(r,t)$ totally undetermined:
$$\frac{\partial}{\partial
\varphi}\left(\frac{\partial}{\partial \theta} -
\cot\theta\right)Y^m_1 \equiv 0 \Longrightarrow h_{\theta\varphi}
= h_{\varphi\theta} = 0\;.
$$
This gives us an additional degree of freedom while making the gauge
transformation which can be utilize to impose the condition $G = K$
to annul $h_{\theta\theta}$ and $h_{\varphi\varphi}$ components:
$$h_{aa} = g_{aa}[K -G]Y^m_1 \Longrightarrow K_1 = K - G = 0,
\,\,\,\,\,\,\,\,\,\,\,\,\,\,\mbox{for $G = K$}\;.
$$
Thus in the canonical gauge there are only three radial functions,
namely, $H_0$, $H_1$ and $H_2$ that need to be determined. From
$\delta R_{tr}$ one can infer that $$H_1 = \im \w r H_2\,,$$ for $l
= 1$. Combining this with eq. (\ref{R_0_2_even}) gives $H_2$,
\bea H_2(r) = \frac{\kappa}{r^2\sqrt{1 + r^2/\ell^2}}\;, \eea
where $\kappa$ is a constant need to be determined according to
the boundary conditions. It can be clearly seen that as $r
\rightarrow \infty$, $H_2(r)\propto (1/r^3)$ approaches zero
($h_{11} \propto 1/r^4\rightarrow 0$). At the origin $H_2$
($h_{11}$) blows up and thus $\kappa$ is set to zero. So for $l =
1$,
\bea H_2 = 0 = H_1 \,.\eea
Now it is easy to find $H_0$ from the remaining equations:
\bea H_0(r) = \frac{c r}{\sqrt{1 + r^2/\ell^2}} \;.\eea
As one can observe $H_0$ approaches the constant $c$ at infinity
(whereas $h_{00} \propto r \rightarrow \infty$) and falls to zero at
the origin. The best choice for $c$ is then to set it to zero and
hence $H_2 = 0$. So in short,
\bea [h_{\mu\nu}^{l = 1, m}]^{(even)} = 0 \longrightarrow
[h_{\mu\nu}^{1m}]_{AdS_4}= 0 \,, \eea
which means there is no electric dipole as well as monopole
perturbation around $AdS_4$ background and all radiative modes have
$l \geq 2$.

{\bf III. Case $l \geq 2$:}

Specializing to the $AdS_4$ background, the case $l > 1$ corresponds
to seven radial functions, which in the canonical gauge, reduce to
only four, i.e., $H_0$, $H_1$, $H_2$, and $K$. In this case $\delta
R_{\mu\nu} + \L_4 h_{\mu\nu}= 0$ implies all the bracket factors in
front of the angular coefficients vanish. The field equation $\delta
R_{\theta\theta} + \L_4 h_{\theta\theta} = 0$ (or $\delta
R_{\varphi\varphi}+ \L_4 h_{\varphi\varphi} = 0$) yields the
relation
\bea H_0 = H_2 \equiv H \,. \eea

Eliminating $H_0$ from the other six equations gives the following
system of radial equations:
\bea (t,r) &:& \im \w\,\left(K^\prime - \fr12\,KA^\prime + \fr1{r}K
-
\fr1{r}\,H\right) + \fr{L(L+1)}{2r^2}H_1 = 0\,,\label{r01}\\[0.6cm]
(t,\theta) &:& \im\w\,(K + H) +
A^\prime\,H_1\,\rme^{A} + H_1^\prime\,\rme^{A} =0\,,\label{r02}\\[0.6cm]
(r,\theta) &:& \im \,\w\,H_1\,\rme^{-A} +
H^\prime-K^\prime + A^\prime H = 0\,,\label{r12}\\[0.6cm]
(t,t) &:& \fr12\,\w^2 (H + 2K) -
\im\w\,\rme^{A}\,\left(\fr{A^\prime}{2} +\fr{2}{r}
+\fr{d}{dr}\right)H_1 -
\fr12\,H^{\prime\prime}\nn\\
&& + \rme^{2A}\,\left(\fr{A^\prime}{2} +\fr1r\right)H^\prime
+\fr12\,\rme^{2A}\,A^\prime K^\prime +
\rme^{A}\left[\fr{L(L+1)}{2r^2} -\fr{3}{\ell^2}\right]H
= 0\,,\label{r00}\\[0.6cm]
(r,r) &:&  \im\w\,\rme^{-A}\left(H^\prime_1+
\fr{A^\prime}{2}H_1\right)
- \rme^{2A}\,\fr{\w^2}{2}\,H \nn\\
&& + \fr{H^{\prime\prime}}{2} - K^{\prime\prime} +
A^\prime\,H^\prime + \fr1{r}\,H^\prime -\left(\fr{A^\prime}{2} +
\fr2{r}\right)\,K^\prime + \rme^{-A}\left[\fr{L(L+1)}{2r^2} +
\fr{3}{\ell^2}\right]
= 0\,,\label{r11}\\[0.6cm]
(\theta,\theta) &:&-\fr{\w^2r^2}{2}\rme^{-A}K + \im\w\,r\,H_1 -
\fr12\,\rme^{A}r^2\,K^{\prime\prime} + r\rme^{A}\,H^\prime\nn\\
&& - \rme^{A}\,\left(\fr{r^2A^\prime}{2}+2r\right)\,K^\prime +
\rme^{A}\,\left(rA^\prime + 1\right)\,H + \rme^{A}\,\left[
\fr{L(L+1)}{2}\,\rme^{-A} - \fr3{\ell^2}\,\rme^{-A} -1 - r
A^\prime\right]K= 0 \label{r22}\,. \eea

Notice that from Eqs.(\ref{r01}), (\ref{r02}) and (\ref{r12}) one
can derive any of the second order equations (\ref{r00}),
(\ref{r11}) and (\ref{r22}) provided the following algebraic
equation
\bea
{\cal G}(r) &=&\left[-2\,\rme^{A} + L(L+1)+rA^\prime\right]\,H +
\fr{\im}{2\w}
\left[-4rw^2 + \rme^{A}\,L(L+1)A^\prime \right]\,H_1\nn\\
&&+\left[ -\frac{6r^2}{\ell^2} + 2\,\rme^{A} - L(L+1)+
   2r^2\w^2\,\rme^{-A} + \rme^{A}\,rA^\prime+
   \rme^A\,\fr{r^2A^{\prime\;2}}{2}\right]\,K= 0
\label{algebraic} \eea
Thus the original system of radial functions has been reduced to a
system of two first-order equations for two unknown functions.  We
can proceed to write this as a single second-order equation
following \cite{Zerilli:1970} by introducing new function ${\cal
E}(r)$ through the following relations
\bea K = \fr{l(l+1)}{r} {\cal E} + \fr{d {\cal E}}{d
r^*}\,,\label{K_even} \eea
\bea H_1 = - i\omega \fr{\ell^2}{\ell^2 + r^2} {\cal E} - i\omega
\fr{\ell^2 r}{\ell^2 + r^2}\fr{d {\cal E}}{d r^*}\,.
\label{H1_even}\eea
So now the Einstein equations for the perturbed metric
(\ref{even_general_metric}) leads to the following equation
\bea \frac{d^2 {\cal E}}{d r^{*2}} + [\w^2 - V_{\rm
eff}^{even}(r)]{\cal E} = 0 \,,\label{even_schrod} \eea
where,
\bea dr^* = \rme^{-A}\,dr\,, \eea
with
\bea V_{\rm eff}^{even} = \left( 1 +
\fr{r^2}{\ell^2}\right)\fr{l(l + 1)}{r^2}\,. \label{P_even} \eea
Notice that the effective potential in both the odd and even cases
are identical unlike the case of pure Schwarzschild
\cite{Zerilli:1970} or Schwarzschild-AdS$_4$ \cite{Card_Lemo:2001}
cases.  This should not surprise us as in the absence of any
gravitational waves scatterer (namely massive source) there should
be no difference between odd and even modes.  Note that the only
invariant of the group is the Laplace-Beltrami operator which is the
wave equation for the free field.  So, in this case we could not
have two wave equations due the symmetry of the AdS$_4$ group.

\section{Stability Analysis}
In order to look at the stability of the modes we should solve
Eqs. (\ref{odd_schrod}) and (\ref{even_schrod}).  For that let's
first write these Schr\"odinger-type equations in the following
form
\bea \rme^{2A} \frac{d^2 \Psi(r)}{d r^2} + A' \rme^{2A} \frac{d
\Psi(r)}{d r} + \rme^A\left[\frac{\w^2 r^2 \rme^{-A} - l(l +
1)}{r^2}\right] \Psi(r) = 0 \,. \label{schro-r}\eea
where $\Psi(r)$ is a generic form for ${\cal M}(r)$ or ${\cal
E}(r)$.  There are two conditions for stability which have to be
satisfied at once: (1) The frequency of the modes has to be
non-positive imaginary and (2) the modes ($\Psi$ here) should
satisfy the completeness condition, i.e., the norm of the solution
to Eq. (\ref{schro-r}) should be finite. Now to solve Eq.
(\ref{schro-r}) we have to deal with the boundary conditions of
the AdS, particularly the one at infinity.  As we have mentioned
earlier there are problems associated with the non-hyperbolicity
of the Anti-de Sitter spacetime. There have been many attempts to
address the ambiguity of the AdS boundary condition at infinity
\cite{AIS:1978, Abbott_Deser:1982, Br_Freed:1982, Burge_Lut:1985}.
We shall adopt the boundary condition introduced in
\cite{AIS:1978} in which the mode function dies off at infinity.
We also demand that $\Psi$ to be well-defined everywhere including
at the origin.

Defining a dimensionless quantity $a$ such that
\bea \ell^2 a = r^2 \rme^{-A} = \frac{r^2}{1 + r^2/\ell^2}\,,\eea
will change (\ref{schro-r}) to the following dimensionless second
order equation
\bea 4 a ( 1 - a) \frac{d^2 \Psi(a)}{d a^2} + 2 (1 - 2 a) \frac{d
\Psi(a)}{d a} + \frac{\w^2 \ell^2 a - l (l + 1)}{a} \Psi(a) = 0
\label{shro_a} \,.\eea
Note that the range of $a$ is finite, i.e. $a \in [0,1]$ ($a$ is
like the normalized $r^*$). We can hope to obtain a hypergeometric
differential equation by rescaling $\psi(a)$
\bea \Psi (a) = \sqrt{a - 1}\, a^{(l + 1)/2}\Phi(a)\,,
\label{rescale_wave}\eea
with
\bea \frac{d \Psi}{d a} & = & \sqrt{a - 1}\, a^{(l + 1)/2}
\left\{\frac{d \Phi}{d a} + \frac{1}{2}\left[\frac{ x + (l + 1)(x
- 1)}{x(x - 1)}\right]\Phi\right\} \,,\\
\frac{d^2 \Psi}{d a^2} & = & \sqrt{a - 1} \,a^{(l + 1)/2}
\left\{\frac{d^2 \Phi}{d a^2} + \left[\frac{ x + (l + 1)(x -
1)}{x(x - 1)}\right]\frac{d \Phi}{d a} + \left[\frac{(l^2 - 1)}{4
a^2} + \frac{l + 1}{2 a (a -1)} - \frac{1}{4 (a -1)^2}\right]\Phi
\right\} \,.\eea
Therefore if $\Phi(a)$ is well-defined everywhere in the range of
$[0,1]$ then (\ref{rescale_wave}) guaranties that $\psi(a)$ is
zero at both the origin and the infinity of the AdS$_4$.
Substituting (\ref{rescale_wave}) into (\ref{shro_a}) yields
\bea 4 a ( 1 - a) \frac{d^2 \Phi(a)}{d a^2} + [6 - 4 l (a -1) - 12
a] \frac{d \Phi(a)}{d a} + [\w^2 \ell^2 a - l^2 - 4 (l + 1)]
\Psi(a) = 0 \label{shro_a2} \,,\eea
which has a solution of the form
\bea \Phi(a) = C_1\, _2F_1\left[1 + \frac{l}{2} - \frac{\w
\ell}{2}, 1 + \frac{l}{2} + \frac{\w \ell}{2}; \frac{3}{2} + l;
a\right] + C_2\,_2F_1\left[1 - \frac{l}{2} - \frac{\w \ell}{2},
\frac{1}{2} - \frac{l}{2} + \frac{\w \ell}{2}; \frac{1}{2} - l;
a\right](-a)^{-(1 + 2 l)/2}\label{shro_solution}, \eea
where $C_1$ and $C_2$ are constant numbers.  In order to have a
well-behavior $\Phi(a)$ we need to set $C_2$ to zero and the first
argument of the hypergeometric function, $_2F_1$, to a negative
(real) integer number such that $\Phi(a)$ becomes a polynomial of
order $N$ at $a \rightarrow 1$ ($r \rightarrow \infty$) , i.e.,
$\Phi(a) \sim a^N \sim a^{-(1 + l/2 - w\ell/2)}$.  This requires
\bea \w \ell = l + 2 (N + 1)\,, \,\,\,\,\,\,\, \mbox{for $l \geq
2$}\,, \label{ads_frequency}\eea
where $N$ is a non-negative integer number.  Thus there is a
constraint on frequency of the gravitational modes in AdS$_4$.
Frequency is quantized and it is always positive-definite (and real)
which means there could be no unstable modes and hence the AdS$_4$
spacetime is stable against small gravitational fluctuations as was.
Equation (\ref{ads_frequency}) also shows that there is gap in
frequency (energy), i.e., it could not ever be zero.  Also note that
$_2F_1$ is complete by construction which means that all the modes
are normalizable.  This complete our proof of stability of AdS$_4$
against gravitational perturbation.  Note that according to
\cite{Burge_Lut:1985}, for the massless minimally coupled scalar
perturbation, the frequency of the modes quantized as $l + 2N + 3$.

\section{Discussion}
In this paper we analyzed  the perturbations on the Anti-de Sitter
spacetime in details. We have shown that a perturbed AdS does not
admit any non-radiative modes while the final form of the
governing equations for radiative odd and even modes are
identical. In proof of stability we explicitly proved the
necessary and sufficient conditions of stability: non-existence of
(positive) imaginary frequency and completeness of the modes. We
showed that the frequency of the gravitational modes must be
quantized.

Here we would like to emphasize on two implicit features in the
discussion of the tensor perturbation of the AdS.  First, note that
Eq. (\ref{schro-r}) is a tensor equation though it may look like the
equation of motion for a scalar, $\Psi$.  In our notation $\Psi$ (or
$\Phi$) corresponds to covariant components of the fluctuations.
There is no rule to exclude other cases and it seems that is more
matter of taste rather than deep physical notion. Second, we
selected a boundary condition that Ref. \cite{AIS:1978} has adopted
for a scalar perturbation at spatial infinity.  This choice of the
boundary conditions at spatial infinity by no means is
gauge-invariant.  What one can do at best is to fix a gauge such
that all the modes die off at spatial infinity. Although one may
argue that supersymmetric boundary conditions will take care of this
problem but as far as the authors are aware there are some
ambiguities associated with the gauge-invariant description of the
boundary conditions at spatial infinity due to the lack of
hyperbolicity of the AdS which results in propagation of information
from and into the spatial boundary in a finite coordinate time
\cite{aadp:2005}.  One way of by passing the absence of any
gauge-invariant boundary conditions in the AdS spacetime is to work
with gauge-invariant variable quantities rather than working in a
particular gauge.  This issue is under investigation and we hope to
address this in our future publication \cite{nayeri:2005}.

At the end we would like to mention that despite the complicated
nature of Regge-Wheeler decomposition one may apply it to higher
dimensional AdS and the cases with brane-worlds.  The latter may
be achieved by considering perturbation  like $\d G_{\mu\nu} =
\kappa^2 \d T_{\mu\nu} = \lambda \d (r - r_0)\d g_{\mu\nu}$, where
$\lambda$ is a constant and $r_0$ is the position of the
brane-world.
\begin{acknowledgments}
We would like to thank Alan H. Guth for many useful discussions and
early collaboration on this project. We also wish to thank S.
Detwhiler, J. Ipser, J. Polchinski, E. Witten and R. Woodard for
many useful discussion.  A.N. also would like to thank MIT Center
for Theoretical Physics where this project was started and mostly
completed.

\end{acknowledgments}

\end{document}